\documentclass[lettersize,journal]{IEEEtran}
\usepackage{balance}
\usepackage{amsmath,amsfonts}
\usepackage[caption=false,font=normalsize,labelfont=sf,textfont=sf]{subfig}
\usepackage{graphicx}
\usepackage{mathtools}

\DeclareMathOperator*{\argmin}{arg\,min}

\newtheorem{theorem}{Theorem}

\begin{document}
\title{Experimental Safe Extremum Seeking\\ for  Accelerators}
\author{Alan Williams, Alexander Scheinker, En-Chuan Huang, Charles Taylor, Miroslav Krstic
\thanks{This work was supported by the Los Alamos National Laboratory, Los Alamos, NM, USA, through the Laboratory Directed Research and Development Project, under Grant  20220074DR.

Alan Williams is with the University of California - San Diego, 92093 San Diego, USA, and Accelerator Operation Technology - Applied Electrodynamics (AOT-AE) at Los Alamos National Lab, 87544 Los Alamos, USA (e-mail: awilliams@lanl.gov).

Miroslav Krstic is with the University of California - San Diego, 92093 San Diego, USA (e-mail: krstic@ucsd.edu).

Alexander Scheinker, En-Chuan Huang, and Charles Taylor are with Los Alamos National Lab (AOT-AE), 87544 Los Alamos, USA (e-mail: ascheink@lanl.gov; 
en-chuan@lanl.gov; cetaylor@lanl.gov).}}

\maketitle

\begin{abstract}
   We demonstrate the recent designs of Safe Extremum Seeking (Safe ES) on the 1 kilometer-long charged particle accelerator at the Los Alamos Neutron Science Center (LANSCE). Safe ES is a modification of ES which, in addition to minimizing an analytically unknown cost, also employs a safety filter based on an analytically unknown control barrier function (CBF) safety metric. 
   
   Accelerator tuning is necessitated by the accelerators being large, with many drifting parameters due to thermal effects and degradation. At the same time, safe operation (the maintenance of state constraints) is  crucial, as damage brings astronomical costs, both financially and in operation downtime. 
   
   Our measured (but analytically unknown) safety metric is the beam current. We perform multivariable Safe ES on three accelerator applications, in which we adapt 4, 6, and 3 magnet strength parameters, respectively. Two of the three applications are for validated simulation models of beamlines at LANSCE: the first for the Proton Radiography (pRad) beamline of 800 MeV protons for spot size tuning; the second on a high performance code, HPSim, for tuning the low energy beam transport (LEBT) region of of 750 keV protons. The third is an experimental tuning of the steering magnets in the LEBT at LANSCE. 
\end{abstract}

\section{INTRODUCTION}
At Los Alamos Neutron Science Center (LANSCE) the 800-MeV proton linear accelerator requires weeks of tuning every year during the maintenance period, as well as hundreds of hours of tuning during operation, due to unknown drifting of components along the approximately 1 km beamline. It is often not possible to use a simulation tool or a model to tune the accelerator offline, due to the complexity of the system and because the system changes with time. Therefore, there is need for real-time optimization of parameters such as magnet strengths, radio-frequency (RF) cavity phases, RF cavity amplitudes, steering devices, etc. to correct the beam towards optimal performance. In this paper we  consider beam loss to be our measure of safety, as it arises often in scenarios relating to machine safety. Furthermore, it can easily be measured in the form of beam current. One may also consider other accelerator applications with a different measure of safety (or unknown constraint) like power draw or radiation level. 

In this work, we use a novel modification of Extremum Seeking (ES) called Safe Extremum Seeking (Safe ES) to solve the problem of \textit{safe optimization} or \textit{constrained optimization}. The motivation for our use of Safe ES is the combination of complexity and lack of diagnostics in challenging charged particle beam tuning tasks in high energy accelerators. The need for a safe tuning algorithm is clearly demonstrated by the beam power of the LANSCE accelerator, which reaches 800 kW, a factor of 80 greater than a typical welding torch. Such a powerful charged particle beam, if not safely and carefully controlled, can instantly burn a hole in the beam pipe of a particle accelerator, destroying the high vacuum system and irradiating nearby components. 

In this work we focus on two very important sections of the LANSCE accelerator, low energy beam transport (LEBT) near the front end of the accelerator and the Proton Radiography facility (pRad) experimental beam line at the very end of the accelerator.

Almost all large particle accelerators have LEBT sections in which the phase space (the collection of positions and velocities of all charged particles) of a relatively low energy beam is first shaped and refined before it is accelerated to high energies in subsequent acceleration sections. The LEBT sections of accelerators are some of the most difficult to tune and control because low energy beam dynamics are dominated by complex collective effects such as space charge forces, which become much less relevant as the beams are accelerated to highly relativistic energies.

Our in-hardware demonstration of the Safe ES approach takes place in the LEBT section of the LANSCE accelerator. The LEBT is directly after the accelerator's beam source and transforms the 750 keV $H^-$ ion beam from a continuous stream of particles into a roughly 600 $\mu$s long beam of individual bunches of ions that are each separated by approximately 5 ns. The bunched beam then enters the first resonant structures of the linear accelerator with the $\sim$5 ns bunch-to-bunch spacing matching the period of the 201.25 MHz resonant electromagnetic fields for subsequent acceleration up to 800 MeV. Tuning of the LEBT region is crucial for LANSCE operations because it sets the initial conditions of the beam that define the rest of the beam transport. Tuning in the LEBT is also challenging because of a lack of diagnostics and because the beam has very low kinetic energy and is very space charge dominated, resulting in a halo of particles around the beam which intercept the beam pipes and accelerator components. We demonstrate experimentally that multi-variable Safe extremum seeking, adapting 3 steering magnet strengths, proved useful in recovering the safe operating condition of the LEBT region.

\begin{figure*}[tb]
\centering
\includegraphics[width=\textwidth]{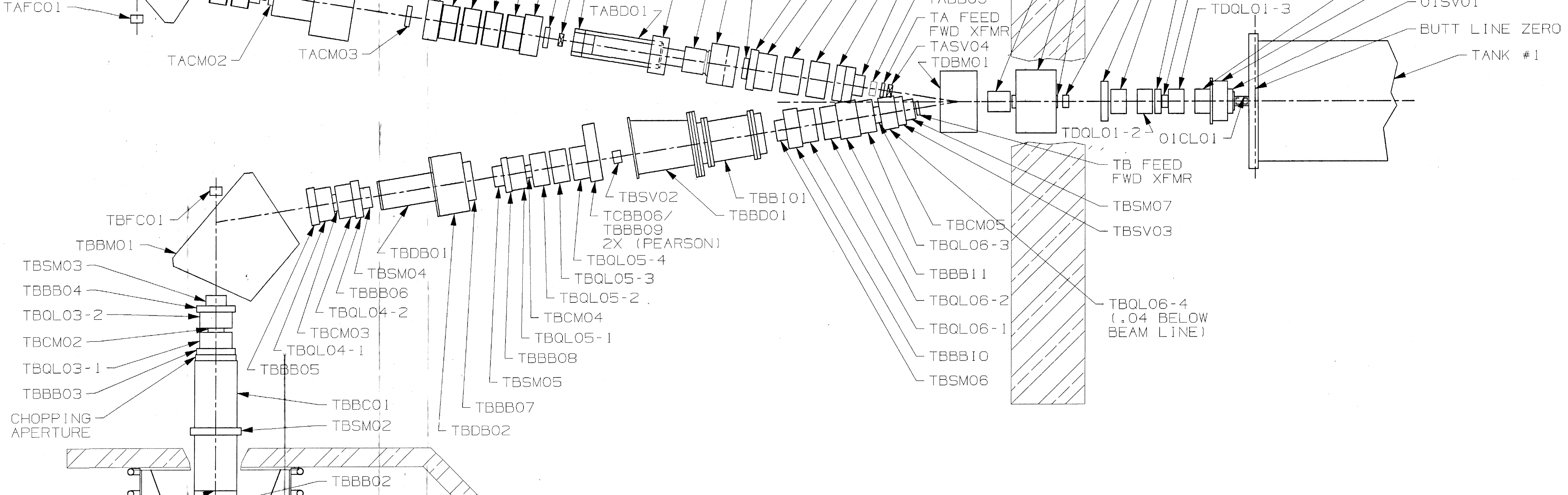}
\caption{A diagram of Line B of the LEBT containing many components including current monitors, bending magnets, steering magnets, focusing magnets, bunchers, and more. Line B delivers $H^{-}$ beam to the first tank of the drift tube linac. Line A, shown partially in the top quarter of the image, delivers $H^{+}$. In Sections \ref{sec:sim_lebt} and \ref{sec:exp_lebt} we tune several components in this section of the accelerator.}
\label{fig:lebt_diagram}
\end{figure*}

We also demonstrate the Safe ES approach via a simulation study of the pRAD experimental facility at LANSCE which is used to characterize the behavior of high explosives, materials under high strain rate, and probe the 3-dimensional density of objects \cite{king1999800}. Under normal operation of pRAD, beam operators view CCD cameras which are pointed at phosphor screens inside the beampipe. The screens, located at various points along the beamline, illuminate when impacted by the beam, showing its' spot size. Daily, several magnet strengths must be manually tuned such that the beam is appropriately sized at one or more of the screens. We show how Safe ES can shape the beam quickly and automatically, while remaining safe conditions as defined by low beam loss.

Safe ES, like classical ES, is 1) real-time 2) easily implementable 3) computationally simple. Yet, it can handle an unknown constraint which may be approximately maintained over the course of tuning. It minimizes an unknown objective function $J(\theta)$, over the parameters $\theta$, while also ensuring that some measure of safety $h(\theta)$ is kept positive. Therefore, our algorithm approximately solves the following problem:
\begin{equation}
\min_{\theta(t)}  J(\theta(t)) \text{ s.t. } h(\theta(t)) \geq 0 \text{ for all } t \in [0, \infty) \end{equation}
where $h$ and $J$ are directly measured signals - or are constructed from directly measured signals, but are analytically unknown. We say that we solve the above problem ``approximately" because ES (and Safe ES) fundamentally relies on approximating of the gradient of unknown functions. Violations of the safety of the system may be made arbitrarily small (or zero in some cases) by appropriately choosing design constants. The design of $h$ is then especially useful in this regard as one may design it such that it provides a `soft' limit on the safe signal of interest. For example, if some safe signal $s(t) \in [0,1]$ must never go below $0.5$, then one may conservatively design $h = s(t) - 0.6$ if one is sure that the chosen small perturbation signal and transient of the ES scheme will not cause a deep excursion of $\theta$ (a change in $h$ more than $0.1$) into unsafe territory. For preventing catastrophic failure to a system in practice, one would implement a higher-level controller to shut down the system once $h$ is sufficiently negative. At LANSCE, high level systems like this will kill all beam if radiation levels spike or beam current is lost to a large degree.

We require the optimization to be done such that a trajectory remains practically safe for all time, or during the entire length of the tuning episode. Methods such as particle swam optimization and genetic algorithms cannot be applied to such scenarios unless a modification is made to ensure that the violation of the trajectory into an unsafe region does not occur.

Note, that we can indeed describe particle accelerators as two unknown, time varying static maps - $J(\theta(t), t)$ and $h(\theta(t), t)$. We imagine that for given parameters $\theta$ at time $t$, we instantly receive a reading of both $J$ and $h$. This is true in many cases when tuning accelerators, as we are not able to do any feedback while the beam is traveling near the speed of light through the beampipe. We assume that only after the beam has completed its journey, will we have a measurements of $J$ and $h$, after which time we can perform the update to the parameters $\theta$.

\section{LITERATURE OVERVIEW}
\subsection{Extremum Seeking}
This work is based on \cite{williams2022practically} and more design changes and analysis in \cite{williamsCDC2023}. The theory of this algorithm is predicated on a classical form of ES \cite{krstic2000stability,ariyur2003real} and the underlying design of the dynamics is based on the QP safety formulation \cite{ames2016control}. We have designed this algorithm such that we do not require the metric of safety to be known, but only measured. The analysis of the presented design is based on  \cite{nesic2010unifying}, proceeded by notable papers \cite{tan2005non,tan2006non} proving semi-global practical asymptotic (SPA) stability.

Constrained ES has been proposed in similar settings \cite{guay2015constrained,labar2019constrained}. In \cite{liao2019constrained} ES is blended with ``boundary tracing", using concepts from linear programming. ES based algorithms are presented in \cite{poveda2015shahshahani}, which can handle both known equality and inequality constraints, modeling dynamics in by evolutionary game theory. Extremum seeking has been used not only in accelerator applications recently but more broadly in automotive \cite{wang2022extremum}, games \cite{ye2020extremum}, medical applications \cite{paz2019model,kumar2019extremum}, and more.

ES algorithms are often classified as a form of ``adaptive control'' in the field of control theory, as it is a way to adapt a set of parameters to optimize an unknown, slowly drifting objective function. The concept of a `barrier' \cite{ames2016control} or safe `boundary' used in the field of ``safe control'' is used throughout this paper and can be thought of as the boundary of the `safe set' $\mathcal{C} = \{ \theta \in \mathbb{R}^n: h(\theta) \geq 0 \}$.

\subsection{Accelerator Tuning and Optimization} The Nelder-Mead simplex method \cite{huang2018robust} and ML based methods have been used to tune accelerators \cite{aiba2012ultra}. Extremum seeking has been studied in simulation for accelerator applications \cite{schuster2007beam}, and was first used experimentally in an accelerator tuning problem to minimize an objective function \cite{modelscheinker}. Since then it has been used extensively in several accelerator applications for free electron laser energy maximization \cite{scheinker2019model}, electron beam trajectory control \cite{7859370}, real-time multi-objective optimization \cite{scheinker2020onlineMO}, and for beam loss minimization \cite{scheinker2021extremum}.

Recently, various machine learning (ML)-based methods have been developed for control and optimization of particle accelerator beams \cite{arpaia2021machine}. Bayesian optimization has become a popular tool in tuning and in some cases has been used to design safety aware tuning algorithms \cite{kirschner2022tuning,kirschner2019adaptive,duris2020bayesian}. Bayesian optimization (and methods based on it), unlike ES, constructs a probabilistic estimate of the unknown functions, in the form of a Gaussian Process (GP), and determines a new point to sample based the fitted function. Neural networks (NN) have been used as surrogate models for magnet control \cite{john2021real} and for simulation-based optimization studies \cite{edelen2020machine}. Neural networks are also being used
for uncertainty aware anomaly detection to predict errant beam pulses \cite{blokland2022uncertainty}, as virtual diagnostics for 4D tomographic phase space reconstructions \cite{wolski2022transverse}, for predicting the transverse emittance of space charge dominated beams \cite{mayet2022predicting}, and for high resolution longitudinal phase space virtual diagnostics \cite{zhu2021high}. Neural network-based deep reinforcement learning (RL) methods have been used for accelerator control \cite{hirlaender2020model}, and in a sample efficient manner, which trains a policy based on data at two beam lines at CERN \cite{kain2020sample}.

Although many ML-based tools have been developed they all suffer major limitations when it comes to time-varying systems. If a system changes then NN, GP, and RL methods all require new data for re-training of their models in order to be applicable for accelerator control. This major limitation is overcome by adaptive model-independent methods such as ES. Adaptive machine learning frameworks combining ES with neural networks have been demonstrated to extend the use of ML for time-varying systems. ES-based adaptive ML has been demonstrated for automatically shaping the longitudinal phase space of short intense electron beams in the LCLS FEL \cite{scheinker2018demonstration} and for creating virtual 6D diagnostics of time-varying charged particle beams \cite{scheinker2021adaptiveML,scheinker2023adaptiveML}.

In all the methods described above, including those using ES, the approach to safety has been some expert-based combination of setting hard bounds on allowed parameter values, adding additional terms to the cost function, and algorithm hyper-parameter tuning. The Safe ES method we demonstrate in this paper reduces the amount of required hyper-parameter tuning and removes the manual design of a tradeoff between safety and optimization which depends on the weights given to safety-related cost function terms relative to objective-related cost function terms. For example, in the real-time multi-objective ES optimization application in \cite{scheinker2020onlineMO}, while the objective was beam spot size minimization, the safety-related term in the cost function was the beam's distance from a desired reference trajectory, and there was a tradeoff between the two depending on the weights. In the approach here, practical safety is always enforced.

\section{ALGORITHM}
\subsection{Algorithm Dynamics}
We will first introduce the set of differential equations describing the multivariate Safe ES algorithm. Given the static maps $J: \mathbb{R}^n \to \mathbb{R}$ and $h: \mathbb{R}^n \to \mathbb{R}$ we define the algorithm dynamics as:
\begin{align}
    \dot{\hat \theta} =& \begin{multlined}[t]
    k \omega_f (-G_J + \\ \min \{G_h^{-2},M^+\} \max\{G_J^T G_h - c \eta_h, 0\} G_h ), \label{eqn:th_dyn} \end{multlined} \\
    \dot G_J =& - \omega_f ( G_J - (J( \hat \theta(t) + S(t) ) - \eta_J)M(t) ), 
    \label{eqn:gj_dyn} \\
    \dot \eta_J =& -  \omega_f ( \eta_J -J( \hat \theta(t) + S(t) )), \label{eqn:etaj_dyn}\\
    \dot G_h =& -  \omega_f (G_h - (h( \hat \theta(t) + S(t) ) - \eta_h)M(t) ), \label{eqn:gh_dyn}\\
    \dot \eta_h =& -  \omega_f ( \eta_h - h( \hat \theta(t) + S(t) )), \label{eqn:etah_dyn} 
\end{align}
where the state variables $\hat \theta, G_J, G_h \in \mathbb{R}^n$ and $\eta_J, \eta_h \in \mathbb{R}$. The overall the dimension of the system is $3n+2$. The maps is evaluated at $\theta$, defined by 
\begin{equation}
    \theta(t) = \hat \theta(t) + S(t) \; .
\end{equation}
The integer $n$ denotes the number of parameters one wishes to optimize over. The design coefficients are $k, c, \delta, \omega_f, M^+\in \mathbb{R}_{>0}$. The perturbation signal $S$ and demodulation signal $M$ are given by  
\begin{align}
    S(t) &= a \left[ \sin(\omega_1 t), \;...\;, \sin(\omega_n t) \right]^T, \\
    M(t) &= \frac{2}{a} \left[  \sin(\omega_1 t), \;...\;, \sin(\omega_n t) \right]^T, \label{eqn:M_def}
\end{align}
and contain additional design parameters $\omega_i, a \in \mathbb{R}_{>0}$ for all $i \in \{1, ..., n\}$.

\subsection{Design}
The tuning parameter vector, $\theta$, is imparted with gradient descent dynamics acting on the objective function \textit{and} gradient ascent dynamics acting on the safety function: 
\begin{align}
    \dot{\theta} =& k \omega_f ( \underbrace{ -G_J }_{ \mathclap{\text{Gradient \textit{Descent} of the Objective \phantom{Explorationnnnnnn}}} } 
    + \overbrace{A G_h)}^{ \mathclap{\text{Gradient \textit{Ascent} of the Safety}} }  + \underbrace{\dot S (t)}_{ \mathclap{\text{Exploration Signal}} } .
\end{align}
The quantity $A$ is a non-negative, state-dependent, scalar function defined as 
\begin{align}
    A &:= \min \{G_h^{-2},M^+\} \max\{G_J^T G_h - c \eta_h, 0\} \geq 0, \\
    & \approx \frac{\max\{\nabla J(\theta)^T \nabla h(\theta) - c h(\theta), 0\}}{\nabla h(\theta)^2} .
\end{align}
It turns ``on/off" to determine whether to consider safety and how much to consider it. Note that $\eta_J, \eta_h, G_J, G_h$ are estimator states which are meant to converge close to the true quantities $J, h, \nabla J, \nabla h$.

To understand the design of $A$, consider the case when the true quantities are known and there is no need for the exploration/perturbation signal. So,

\begin{equation}
    \dot \theta (t) = u \label{eqn:simple_sys},
\end{equation}
for $\theta$ and $u$ in $n$-dimensions. If we consider the gradient descent term $u(\theta) =-\nabla J(\theta)$ to be the nominal controller, we can employ the formulation given by a QP \cite{AmesAutomotive,AmesCruiseControl} to find an additive safety term which, when combined with the nominal, yields an provably safe controller (i.e. the safe set $\{h(\theta) \geq 0 \}$ is said to be positively invariant). The safety aware control law is $u_s = -\nabla J(\theta) + \bar{u}$, and $\bar{u}$ is given by the QP
\begin{align}
    \bar{u} = \argmin_{v \in \mathbb{R}^n} ||v||^2 \quad \text{subject to} \label{eqn:QP1} \\
    c h(\theta) + \nabla h(\theta)^T(-\nabla J(\theta) + v) \geq 0, \label{eqn:QP2}
\end{align}
for some positive constant $c$. It was showed that there is an explicit solution for $\bar{u}$:

\begin{equation}
\bar{u}(\theta) =  \dfrac{\nabla h(\theta)}{||\nabla h(\theta) ||^2} \max\{ \nabla h(\theta)^T \nabla J(\theta) - c h(\theta),0\}. \label{eqn:sf_term_exact}
\end{equation}
It can now be seen that $\bar{u}(\theta) \approx A \nabla h(\theta) \approx A G_h$. 

\subsection{Theory}

For convenience of reader, in this brief section we include theoretical results guaranteed by the safe ES controllers that we implement. These theoretical results are imported from our work \cite{williamsCDC2023}. Their presence in this paper lends some clarity and precision about what we aim to attain in our experimental and simulation tests --- which is (theoretically semi-global) ``practical'' safety and stability. 

Safe ES has a number of desirable properties, the most notable of which is practical safety. Another desirable property is that the control law is smooth, and trajectories avoid crashing into an unsafe region without first trying to skirt around them while they are some distance away. Additionally the formulation of the QP in \eqref{eqn:QP1} - \eqref{eqn:QP2} guarantees that the modification to the nominal control always yields the final control law smaller than the nominal ($u_s \leq u$), for $h>0$. This is a practical benefit for the control designer who must choose the design parameters such that the adaptation dynamics, given in the right-hand side of $\eqref{eqn:th_dyn}$, must be made sufficient small for stability purposes. Yet another benefit is that when the trajectory starts in the unsafe region ($h<0$), the value of $h$ is guaranteed to monotonically increase in time (provided certain assumptions like the gradient of $h$ does not vanish), escaping to more safe parts of the state space.

Suppose $J$ has a unique constrained minimizer on the set $\{ \theta : h(\theta) \geq 0\}$ given by $\theta^*_c$. We define 
\begin{align}
    \tilde \theta &:= \hat \theta - \theta^*_c \label{eqn:tildetheta_def},\\
    z &:= [G_J^T, \eta_J, G_h^T, \eta_h]^T - \mu(\tilde \theta, a) \label{eqn:z_def},\\
    \begin{split} \label{eqn:D_def}
        D(\tilde \theta+ \theta^*_c) &:={} [\nabla J(\tilde \theta+ \theta^*_c)^T, J(\tilde \theta+ \theta^*_c), \\ & \nabla h(\tilde \theta+ \theta^*_c)^T, h(\tilde \theta+ \theta^*_c)]^T ,  
    \end{split} \\    
    \mu(\tilde \theta,a) &:= D(\tilde \theta+ \theta^*_c) + O(a). \label{eqn:mu_def}
\end{align}

The following theorems state the general convergence and safety properties of the Safe ES algorithm implemented in this work. Under mild assumptions on the objective function $J$ and the safety/barrier function $h$ we have shown the following \cite{williamsCDC2023}.

\begin{theorem} [Semi-Global Practical Stability] \label{thm:spa_stable}
Under the assumptions given in \cite{williamsCDC2023}, there exists $\beta_\theta, \beta_\xi \in \mathcal{KL}$ such that: for any positive pair $(\Delta, \nu)$ there exist $M^+, \omega_f^*, a^*>0$, such that for any $\omega_f \in (0, \omega_f^*)$, $a \in (0,a^*)$, there exists $k^*(a)>0$ such that for any $k \in (0, k^*(a))$ the solutions to \eqref{eqn:tildetheta_def}-\eqref{eqn:z_def} satisfy
\begin{align}
    || \tilde \theta(t) || &\leq \beta_\theta \left( ||\tilde \theta (t_0)||, k \cdot \omega_f \cdot (t-t_0) \right) + \nu, \label{eqn:theta_KL_bound} \\
    || z(t) || &\leq \beta_\xi \left( ||z (t_0)||, \omega_f \cdot (t-t_0) \right)  + \nu, \label{eqn:z_KL_bound}
\end{align}
for all $||[\tilde \theta(t_0)^T, z(t_0)^T ]^T|| \leq \Delta$, and all $t \geq t_0\geq 0$.
\end{theorem}

We use the term ``SPA stability'' to refer to the notion of semi-global practical asymptotic stability \cite{tan2005non}. We have also proven a practical result for stability.

\begin{theorem}[Semi-Global Practical Safety] \label{thm:prac_safety}
    Suppose Theorem \ref{thm:spa_stable} holds. For any $\Delta>0$ there exists $\delta^*>0$ such that for any $\delta \in (0, \delta^*)$ : there exists $M^+, a^{**}, \omega_f^{**}>0$ such that for any $a \in (0, a^{**})$, $\omega_f \in (0, \omega_f^{**})$, there exists $k^{**}(a)>0$ such that for any $k \in (0, k^{**}(a))$,
\begin{equation}
    h(\theta(t))  \geq h(\theta(t_0)) e^{-c k \omega_f (t-t_0)} + O(\delta), \label{eqn:practical_safety_inequality}
\end{equation}
     for all $||[\tilde \theta(t_0)^T, z(t_0)^T ]^T|| \leq \Delta$ and for all $t \in [t_0, \infty]$. 
\end{theorem}

This tells us that trajectories that start safe remain arbitrarily safe and trajectories that start unsafe will improve to an arbitrarily safe region. We cannot expect that with a gradient based estimate scheme like ES we will remain or escape to a perfectly safe region - but we can guarantee that a proper choice of $h$ and $k, \omega_f,a$ we can achieve practical safety. As mentioned earlier, this result also implies that for some problems (for a specific $J,h$) designing $h$ with some margin can indeed guarantee perfect safety. The result in Theorem \ref{thm:prac_safety} is also an elegantly analogous to Theorem \ref{thm:spa_stable}, and is therefore referred to as ``practical safety''  in the spirit of the term ``practical stability'' coined in \cite{tan2005non}.

\subsection{Implementation of \eqref{eqn:th_dyn} - \eqref{eqn:M_def}}
To implement the extremum seeking algorithm we integrate the differential equations \eqref{eqn:th_dyn} - \eqref{eqn:etah_dyn} numerically. Expressing the equations using $x^T=[\theta^T, G_J^T, \eta_J, G_h^T, \eta_h]$, we have $\dot x = f(x)$. With some initial condition $x(0) = x_0$, we compute $x_n = x_{n-1} + f(x) dt$ and the parameters $\theta$ can be set with new values iteratively. The value $dt$ is chosen such that period of oscillation $T = 2 \pi / \omega_i \gg dt$. For all implementations of Safe ES shown in this paper, we specify $M^+=10^4$.

Note that unlike the version of ES in \cite{scheinker2014extremum}, we rely on the explicit estimation of gradients. Therefore, in all the results and plots given we ``warm up'' the algorithm first by settings $k=0$ for the first 1-3 periods of the perturbations $S(t)$, before turning on the algorithm and simulating with a nonzero value of $k$. This allows the estimated quantities to converge more closely to their true values before the adaptation of the parameter of $\theta$ begins. This technique can be thought of as a trick to more accurately initialize the estimator states $G_J, \eta_J, G_h,$ and $\eta_h$.

We choose a value of $a$ that causes the oscillatory response in the measurements of $J(\theta(t)), h(\theta(t))$ corresponding to the frequencies $\omega_i$. We set $\omega_f$ roughly the same order of magnitude as $\omega_i$ - a smaller $\omega_f$ can be chosen if one wishes to add more smoothing to the estimator states. Finally, we gradually increase $k$ from zero until sufficient performance is observed $J(\theta(t)), h(\theta(t))$. Choosing $k$ too large may cause instability. 

\section{ACCELERATOR APPLICATIONS}

\subsection{Simulated pRad Tuning}
For experimental studies on the tuning of Line C, which is the section of the accelerator responsible for delivering beam to pRad, we use the particle beam dynamics code TRANSPORT, which has been validated with measurement data and is shown to give accurate predictions of the LANSCE beam profile \cite{Roy_2018}. The code contains relevant quadrupole magnet strengths - which in practice are manually tuned by hand to achieve a spot size required for an optimal delivery of beam to the experiment.

The TRANSPORT model is a beam envelope model, that models the bulk behavior of the beam which is represented as an ellipse in 6 dimensions (3 spatial lengths and 3 spatial velocities). Even though it can model space charge effects, it can not simulate particle loss. Therefore, we introduce a model of loss which is applied after TRANSPORT computes the beam dynamics solutions. This model is not based on data but gives us sufficient complexity to demonstrate our algorithm. We model the percentage of beam remaining $b$ [\%] as a function of the integrated beam size of the TRANSPORT solution:
\begin{multline}
    b =  1 - k_1 \int_{0}^{s_{end}} (X(s) + Y(s)) ds \\ 
    -k_2 \int_{0}^{s_{end}} [((X(s) - r_p)^+)^2 +((Y(s) - r_p)^+)^2 ] ds,
\end{multline}
where $X^+ := \max\{X,0\}$ and $r_p$ is the radius of the beampipe. We choose $k_1<< k_2$ and the equation describes 1) a small linear loss on the size of the beam that always applies at any point $s$ 2) a large quadratic loss applying only when the ellipse makes contact with the beam pipe. The small linear loss models the persistent loss which always occurs due to beam halo, and the large quadratic loss describes the sharp and sudden loss which only occurs when the central mass of the bunch comes close to the beam pipe. This model provides sufficient complexity which captures the dynamics of loss realistically, although it has not been thoroughly validated with data, as the TRANSPORT code itself has, which only describes the dynamics of the beam ellipse. We also provide the LANSCE magnet names used in this study: ``AQM1",``AQM2", ``XQM3", ``XQM4".

The goal for this simulation study relates exactly to a tuning problem which must usually be performed in the control room. We desire to track a specific spot size of beam as a point in the beamline, directly preceding the pRad experimental dome. Operators usually are tasked with this job, but we will show Safe ES is capable of performing the task safely. We will track the size at the end of the simulation $\sigma_x$, $\sigma_y$ [m] and define the objective as 
\begin{equation}
J(t) = (\sigma_x(t) - 0.025)^2 + (\sigma_y(t) - 0.025)^2 ,
\end{equation}
and so the desired spot size is a $2.5$ cm circle. We use values of $k_1 = 0.02$, $k_2 = 100$. We make the simple choice of
\begin{equation}
    h(b(t)) = b(t) - 0.80,
\end{equation}
with no scaling to either $h$ and $J$, which was done earlier in the HPSim study to achieve the same order of magnitude of $J$ and $h$.

\begin{figure}[t]
\centering
  \includegraphics[width=0.47\textwidth]{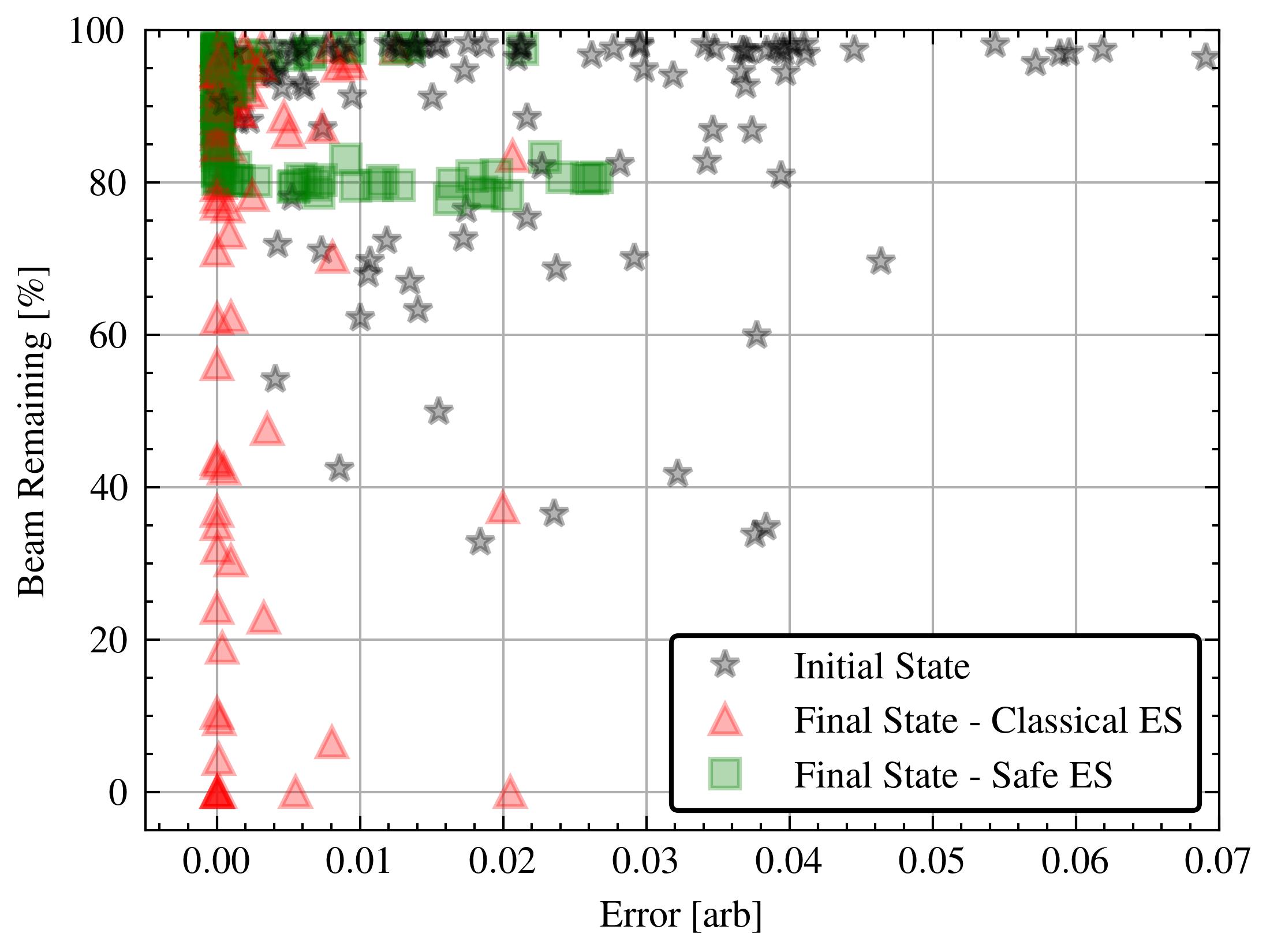}
  \caption{For 100 episodes, we plot both the initial and final error and beam remaining values for safe and unsafe extremum seeking.}
  \label{fig:prad_error_summary}
\end{figure}

\begin{figure}[t]
\centering
  \includegraphics[width=0.47\textwidth]{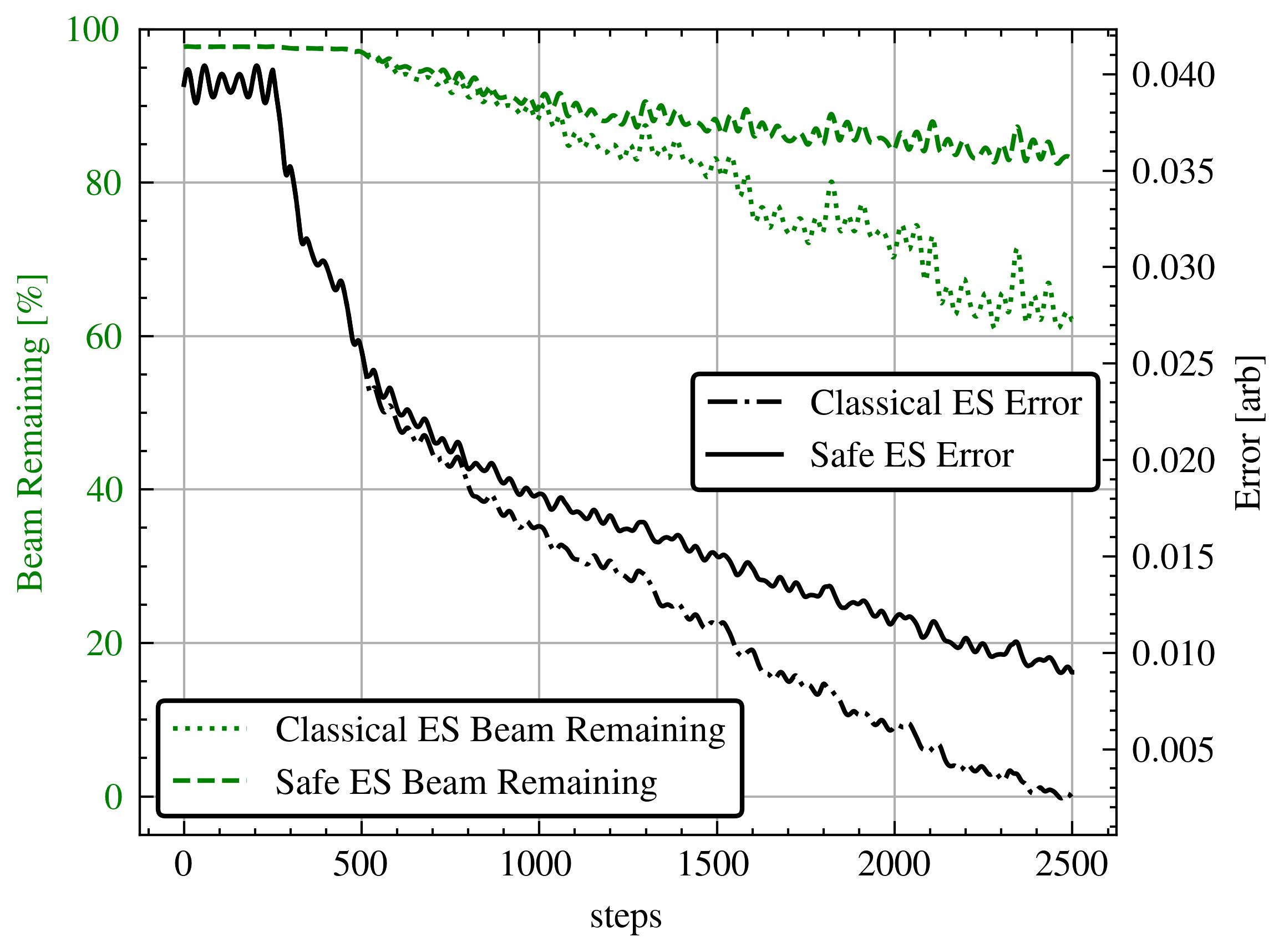}
  \caption{Percentage of beam remaining and error for safe and unsafe extremum seeking, for one given trajectory.}
  \label{fig:prad_error_beamre_summary}
\end{figure}

\begin{figure}[t]
\centering
  \includegraphics[width=0.47\textwidth]{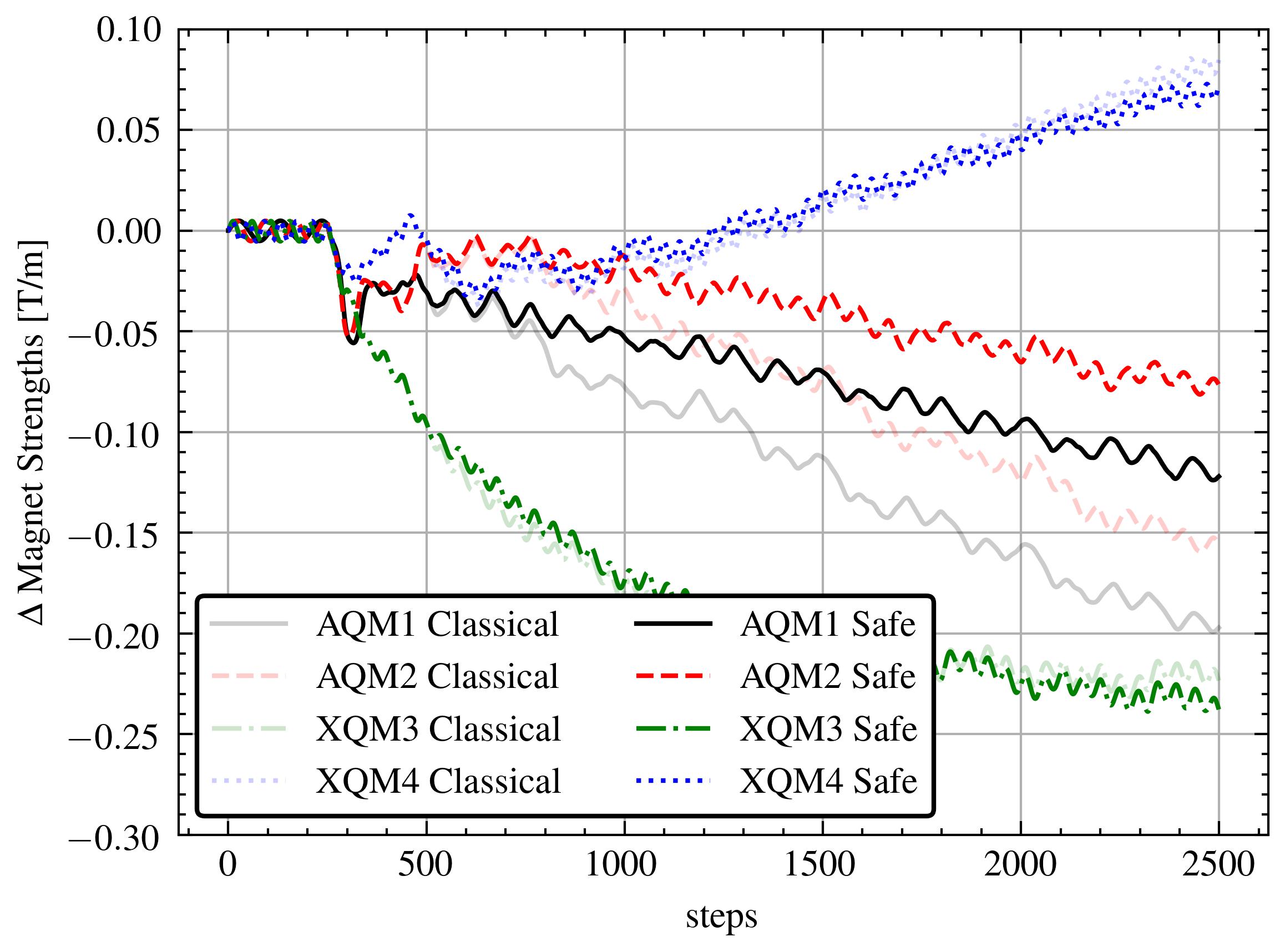}
  \caption{The change in magnet strength trajectories, $\hat \theta - \hat \theta_0$, of Safe ES compared with Classical ES.}
  \label{fig:prad_safe_vs_classical}
\end{figure}

 We randomly initialize the magnet settings at up to $20$\% of their default settings, keeping points which have at least $20$\% beam remaining. This is so that we do not assume we start the simulation in an unrealistic initial condition (catastrophic beam loss) which may crash the simulation. Then we run both Classical ES and Safe ES (with coefficients above) for 2,500 steps and compute the objective error and beam remaining values at the end of the episodes. We do this for 100 different initial conditions and show the initial condition and final condition of both the Safe ES and Classical ES trajectories in Fig. \ref{fig:prad_error_summary}. The Safe ES parameters used are $dt = 0.012$, $a = 0.005$, $k = 0.04$, $\omega_f = 10$, $\omega_1 = 5$, $\omega_2 = 7$, $\omega_3 = 11$, $\omega_4 = 13$ and $c=1$. 

 To construct the Classical ES algorithm, we simply use the dynamics of \eqref{eqn:th_dyn} - \eqref{eqn:etaj_dyn} without the safety term in the dynamics of \eqref{eqn:th_dyn}:
\begin{align}
    \dot{\hat \theta} =& - k \omega_f G_J , \label{eqn:th_dyn_classical} \\
    \dot G_J =& - \omega_f ( G_J - (J( \hat \theta(t) + S(t) ) - \eta_J)M(t) ),  \label{eqn:gj_dyn_classical} \\
    \dot \eta_J =& -  \omega_f ( \eta_J -J( \hat \theta(t) + S(t) )). \label{eqn:etaj_dyn_classical}
\end{align}
We choose the parameters of the Classical ES scheme to be the same as those used in the Safe ES scheme - apart from the irrelevant value $c$ which does not appear in \eqref{eqn:th_dyn_classical} - \eqref{eqn:etaj_dyn_classical}.

Fig. \ref{fig:prad_error_summary} depicts the effectiveness of the algorithm with respect to various initial conditions. We show for a 100 randomly chosen initial conditions, which have small and large values of both performance and safety, all trajectories of Safe ES eventually achieve approximately 20\% or better beam remaining - albeit with various levels of performance. Classical ES shows better performance in general (more red arrows are grouped on the left-hand side of the plot) but various levels of safety.

In Fig. \ref{fig:prad_error_beamre_summary} and \ref{fig:prad_safe_vs_classical} we choose one trajectory of the 100 episodes and plot the magnet trajectories as well as the beam remaining signal and error signal along the trajectory. As expected, the Classical ES trajectory in Fig. \ref{fig:prad_error_beamre_summary} finds itself traveling into an unsafe region in favor of better performance, while the Safe ES trajectory still achieves good performance but does so while keeping the system safe. Running the simulation for longer than 2,500 steps may also indeed show that Safe ES in this case may ultimately achieve near perfect performance by taking a longer path through parameter space.

\subsection{Simulated LEBT Tuning} \label{sec:sim_lebt}
HPSim \cite{pang2015advances} is a GPU accelerator particle tracking code that simulates all of the RF and magnetic devices used at LANSCE and accurately describes the physics of beams, including challenging collective effects like ``space charge''. It has been used to demonstrate tuning algorithms \cite{modelscheinker} and is currently being developed as an online digital twin of the LANSCE linear accelerator (linac) \cite{huang2021enhanced}.

The LEBT section delivers the 750 keV beam from the source to the drift tube linac (DTL) accelerating structures of the linac. In the example presented here, we apply Safe ES to automatically tune 6 of the quadrupole magnets in LEBT which are responsible for confining the beam to the pipe as it is transported. We have specified that we want the smallest possible beam entering the DTL, yet we must attain approximately greater than 80\% of the beam while doing so. Although this is not precisely the goal of the tuning the LEBT beamline, it gives us with a realistic scenario showing the benefits of Safe ES when we are not sure if gradient descent of the objective function leads the parameters in a safe direction or not.

In this simulation study we compare Safe ES with Classical ES with a cost function modified with a small weight on the safety of the system, as this is often a trick used to incorporated safety considerations in the Classical ES scheme. We will also demonstrate the qualitative behavior of the $c$ parameter in the Safe ES design and show that Safe ES in general leads to a configuration of the accelerator which leads to less contact with the beam and sensitive structures in the beampipe.

Given the goals and safety concerns we have described, we choose the following cost function to minimize,
\begin{equation}
    J(t) = w_J(\sigma_x(t)^2 + \sigma_y(t)^2) ,
\end{equation}
and we also choose the function $h$ as
\begin{equation}
    h(t) = w_h (b(t) - 0.80),
\end{equation}
where $b, \sigma_x, \sigma_y$ are signals measured in time and $w_J, w_h$ are chosen such that $J, h$ are approximately on the order of $10$. Using HPSim we determine 6 quadrupoles along the beamline to use for tuning which are roughly equally spaced in $s$ - the variable used to describe the longitudinal positive along the beamline. We provide the LANSCE magnet channel names used in this study: ``TBQL005V01", ``TBQL005V02", ``TBQL005V03", ``TBQL005V04", ``TBQL006V03", ``TBQL006V04". The physical locations of the components can be shown in Fig. \ref{fig:lebt_diagram} as ``TBQL05-1'', ``TBQL05-2'', ..., ``TBQL06-4''.

\begin{figure}[t]
\centering
  \includegraphics[width=0.47\textwidth]{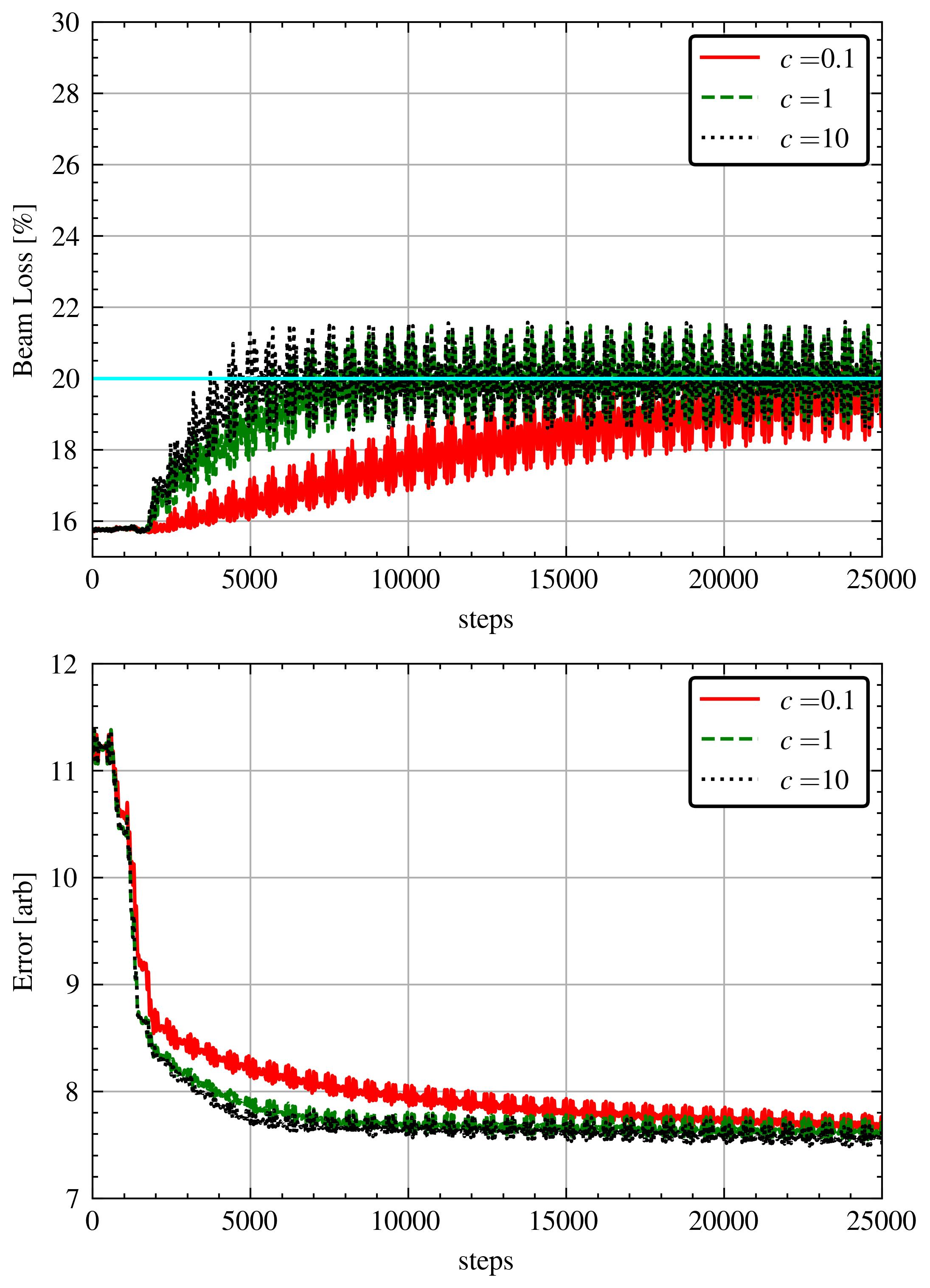}
  \caption{Percentage of beam loss and error for Safe ES with different values of $c$.}
  \label{fig:hpsim_varying_c}
\end{figure}

\begin{figure}[t]
\centering
  \includegraphics[width=0.47\textwidth]{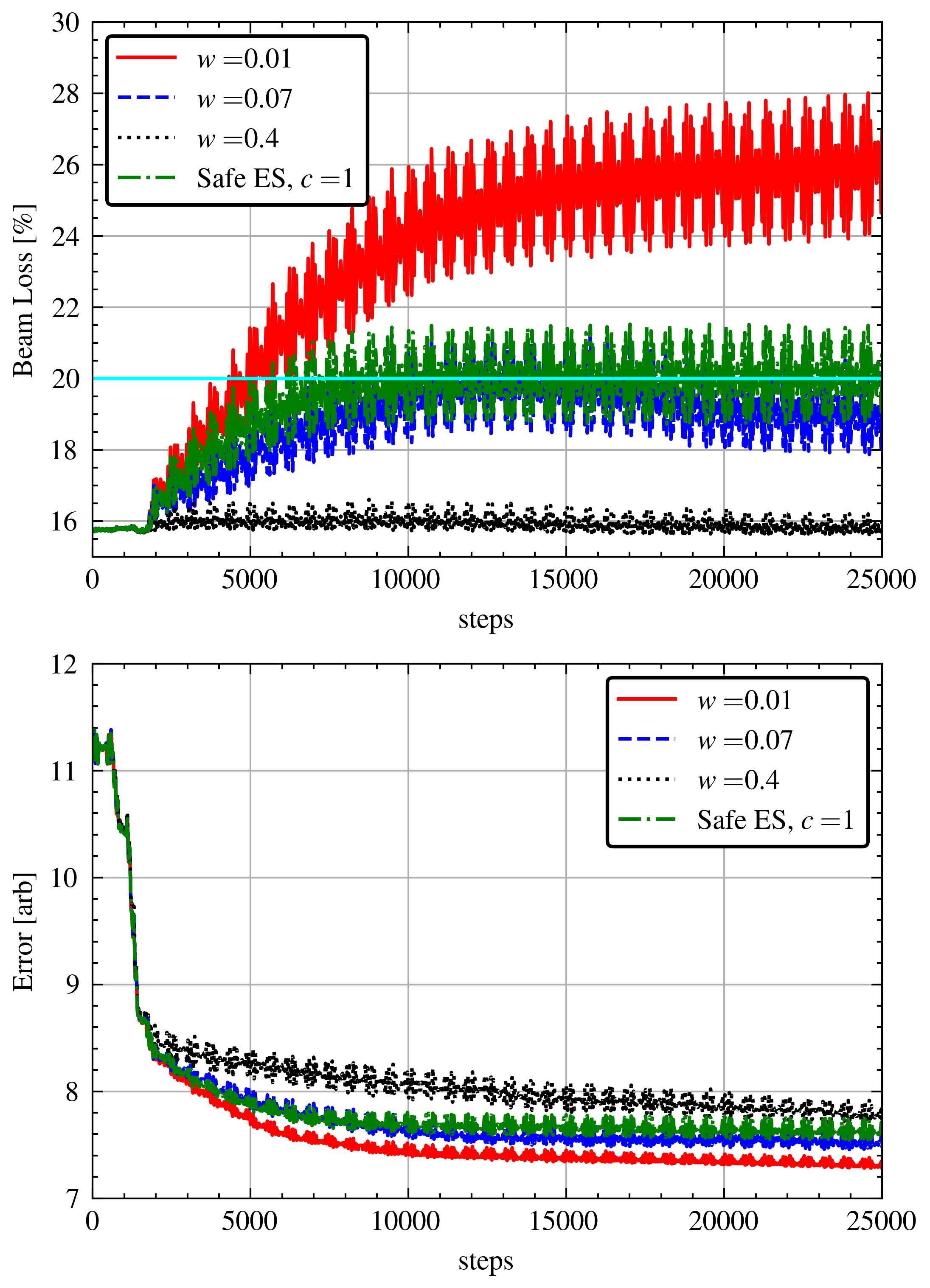}
  \caption{Percentage of beam loss and error for Classical ES with various weights $w$, plotted against one trajectory of Safe ES. Note that the Error values shown in the bottom plot is $w_J(\sigma_x(t)^2 + \sigma_y(t)^2)$, for each of the 4 trajectories given.}
  \label{fig:hpsim_varying_w}
\end{figure}

It is a common strategy to handle the safety problem in Classical ES with a modified cost:
\begin{equation}
    J(t) = w_J(\sigma_x(t)^2 + \sigma_y(t)^2) - w b(t) \label{eqn:mod_cost}
\end{equation}
where $w$ is chosen sufficiently large to (with some luck) yield a safe trajectory. We compare our algorithm with this strategy and show that in this case it is not possible to know exactly how to specify $w$ without achieving some loss in performance or loss of safety. While for any choice of our hyper-parameter $c$ (which is the most analogous hyper parameter to $w$ and in some sense governs the safety-versus-performance tradeoff), we achieve approximate safety in all cases. This is because $c$ does not change the whether or not we maintain approximate safety, but simply how fast the trajectory is allowed to approach the unsafe region.

The Safe ES parameters used are $dt = 0.005$, $a = 0.1$, $k = 0.05$, $\omega_f = 10$, $\omega_1 = 5$, $\omega_2 = 7$, $\omega_3 = 11$, $\omega_4 = 13$, $\omega_5 = 17$, $\omega_6 = 19$. We choose $w_J = 50$ and $w_h = 100$ so that the functions $h$ and $J$ are roughly of the same order. 

The Classical ES algorithm uses the same parameters (apart from $c$) and is implemented in the same way as in the pRad tuning example using \eqref{eqn:th_dyn_classical} - \eqref{eqn:etaj_dyn_classical}. 

\begin{figure}[t]
\centering
  \includegraphics[width=0.47\textwidth]{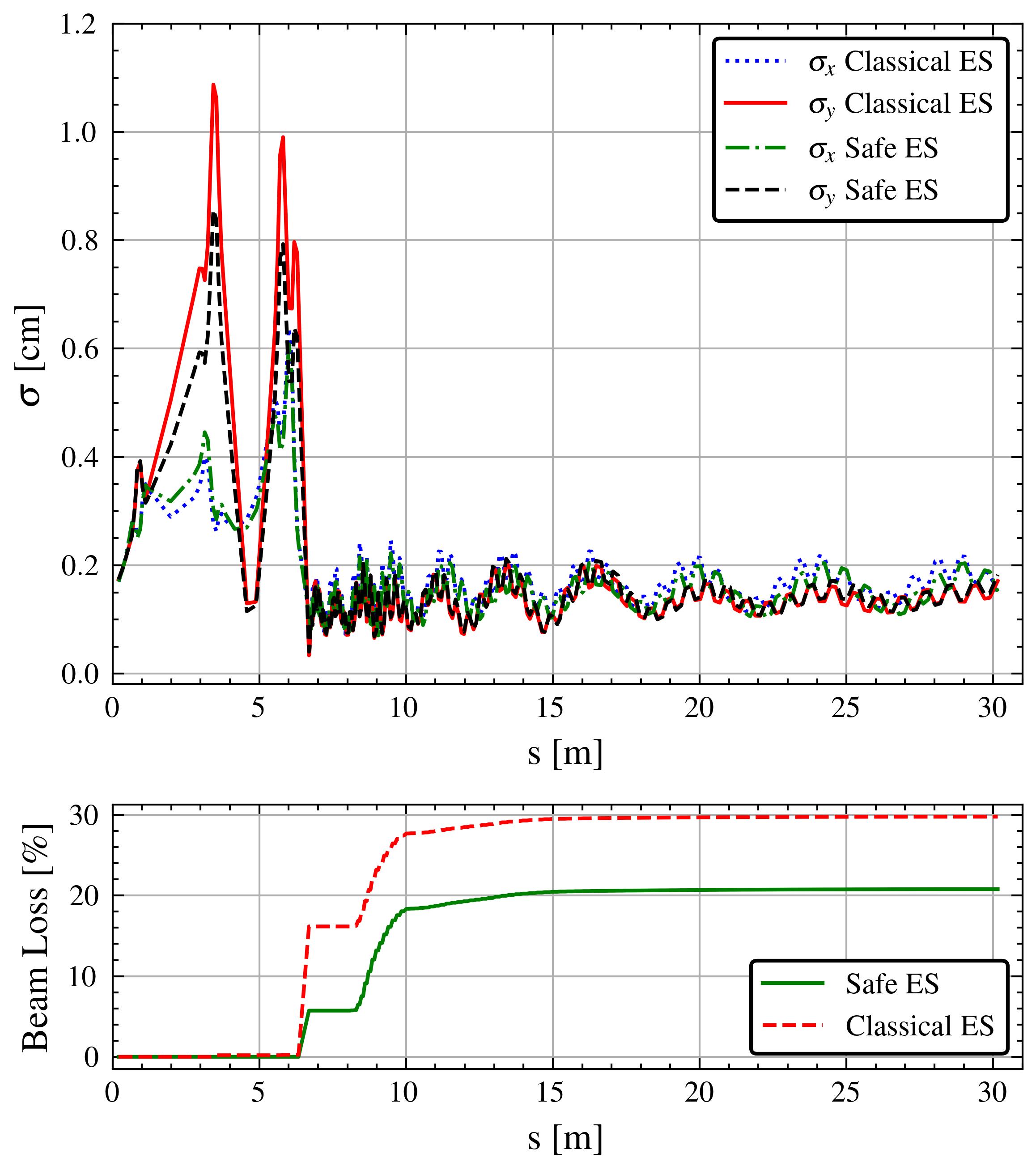}
  \caption{The beam envelope along $s$, depicting the transverse size of the beam throughout the LEBT region and through the first 4 DTL modules in Classical ES with $w=0$ and Safe ES with $c=1$.}
  \label{fig:hpsim_envelope}
\end{figure}

In Fig. \ref{fig:hpsim_varying_c} we show that for various values of $c$, approximate safety is enforced. The smaller values of $c$ dictate a slower rate of approach of the trajectory towards the barrier. Fig. \ref{fig:hpsim_varying_w} shows that Classical ES with a modified cost function, in \eqref{eqn:mod_cost}, does not always remain safe and the weight $w$ cannot be known ahead of time to guarantee best performance and safety. It turns out that choosing $w \approx 0.07$ in the Classical ES scheme would deliver performance and safety comparable to that of the Safe ES controllers, but this choice would require excessive effort, or extra system knowledge, on the part of the control designer.

In Fig. \ref{fig:hpsim_envelope} we compare the beam `envelope' which was found with Safe ES versus Classical ES. The coordinate $s$ runs longitudinally along the beam path, and Fig. \ref{fig:hpsim_envelope} plots the transverse size of the beam, which is the actual height ($\sigma_y$) and width ($\sigma_x$) of the beam if one were to look down the beam pipe. The algorithm has no knowledge of this information, only the measurement of $J$ and $h$. The figure also provides information about at which point beam loss is occurring as the sharp change in the size of the beam marks where the start of the first DTL modules lies. The Classical ES trajectory used was for that of $w=0$ which yielded approximately a final loss value of about 30\%. It was compared to the envelope found using Safe ES with $c=1$. Fig. \ref{fig:hpsim_envelope} demonstrates that the transverse size of the beam along the beam pipe was smaller in the case of Safe ES. This is physically what we expect given that large beam losses are expected to occur when the beam comes in contact with beampipe.

\subsection{Experimental LEBT Tuning} \label{sec:exp_lebt}
\begin{figure}[t]
\centering
  \includegraphics[width=0.47\textwidth]{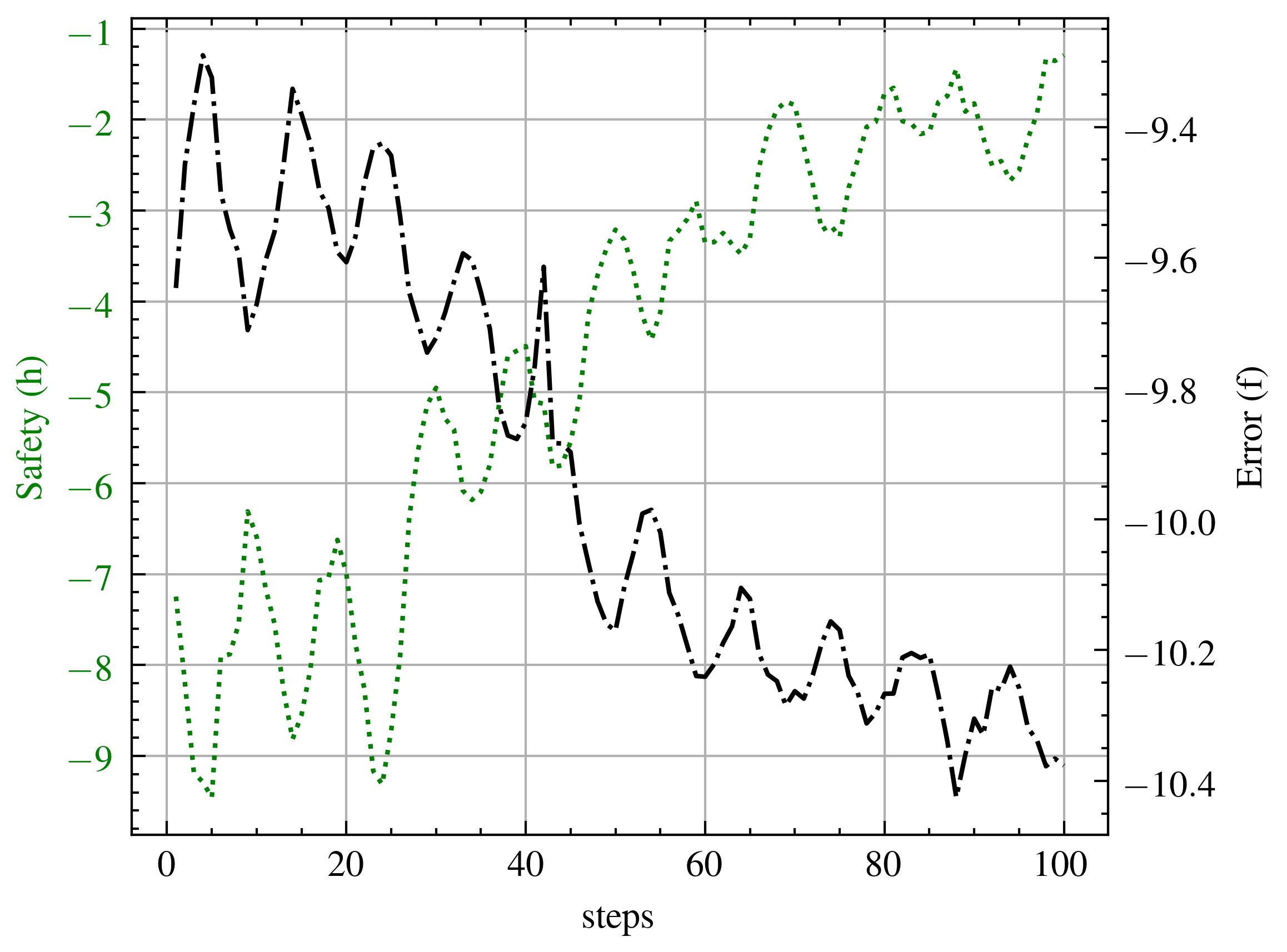}
  \caption{Barrier function and objective function over the first 100 steps using a 10 point average of the noisy measurements of $I_b$ and $I_c$.}
  \label{fig:experimental_safety_and_error}
\end{figure}

\begin{figure}[t]
\centering
  \includegraphics[width=0.47\textwidth]{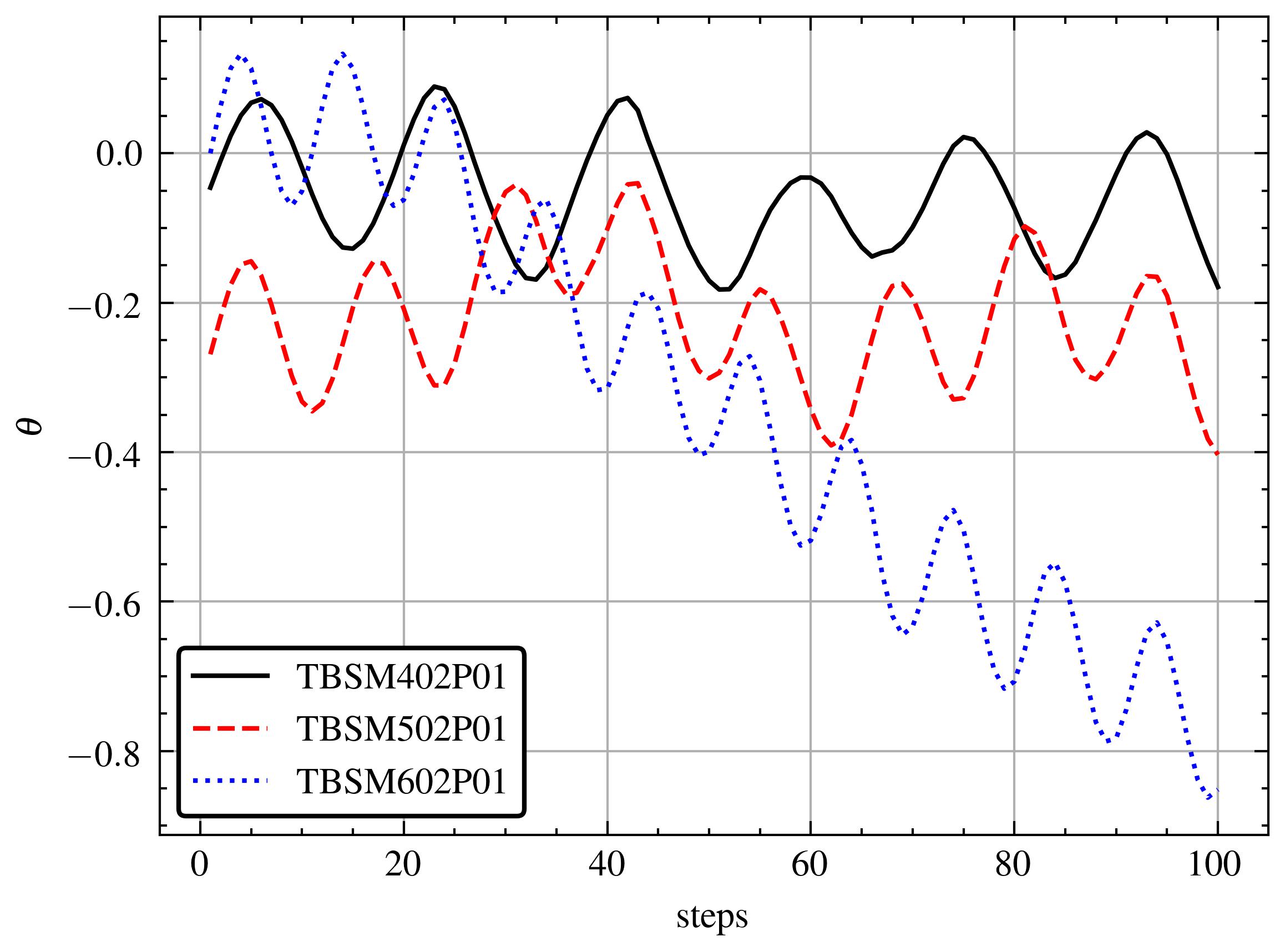}
  \caption{Parameter trajectories over the first 100 steps using a 10 point average of the noisy measurements of $I_b$ and $I_c$.}
  \label{fig:experimental_parameters_beamtime}
\end{figure}

\begin{figure}[t]
\centering
  \includegraphics[width=0.47\textwidth]{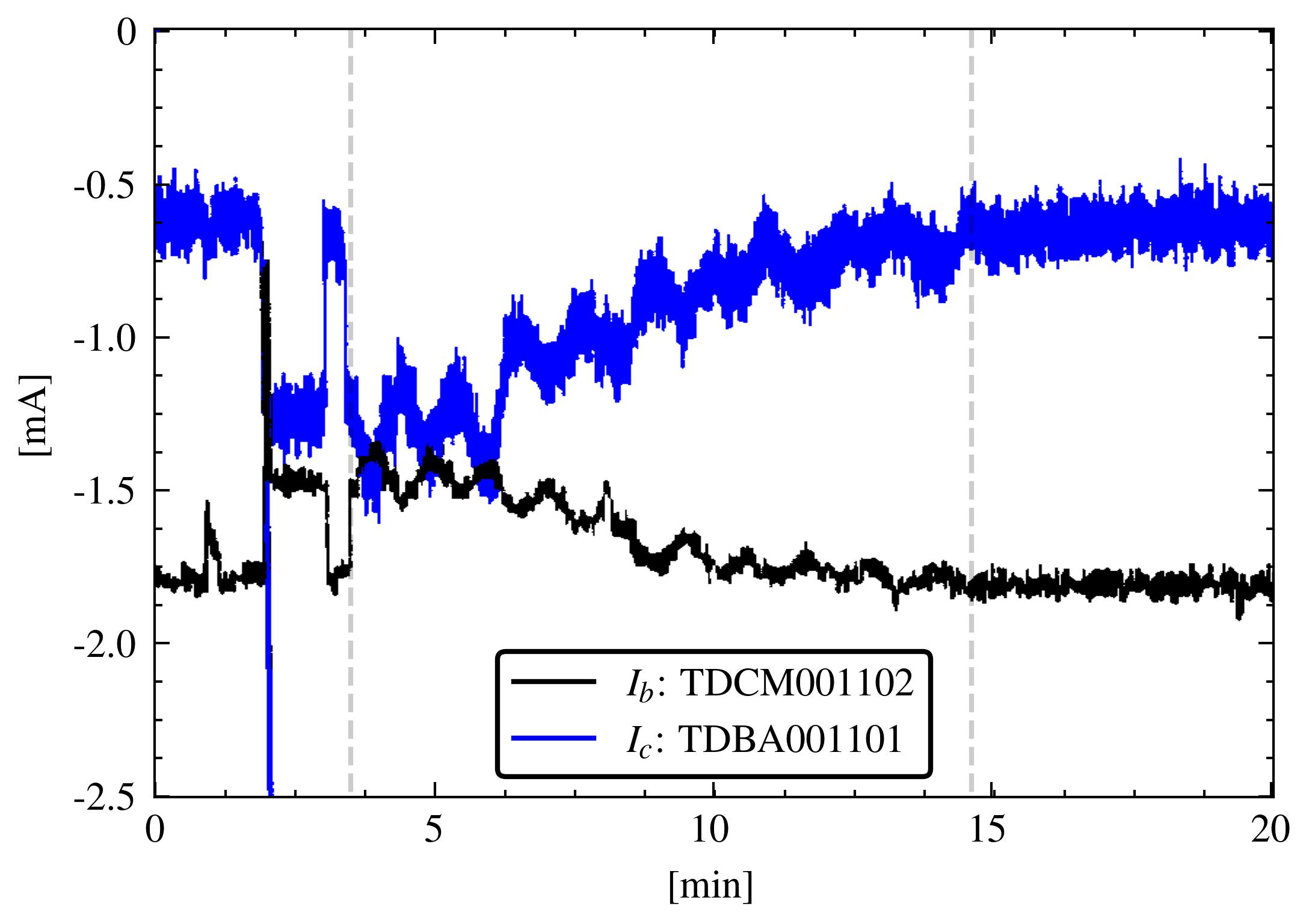}
  \caption{Raw plot of $I_b$ and $I_c$ taken from the LANSCE control room monitor for process variables. The first dotted vertical line marks the start of the algorithm, the second dotted line marks where recordings in Fig. \ref{fig:experimental_safety_and_error} and Fig. \ref{fig:experimental_parameters_beamtime} finish.}
  \label{fig:screenshot_safety_and_error}
\end{figure}
For the in-hardware LEBT demonstration, the ES objective function, $J$, was a measure of beam remaining $I_b$ (negatively proportional to beam loss), while the safety constraint, $h$, was a measure of how much beam is being lost at a collimator in the LEBT, $I_c$. The amount of beam that hits the collimator is proportional to the transverse beam size in this section of the accelerator and it is crucial to keep that beam size small enough to reduce how much beam is intercepted. At the same time, pinching the beam down to too small of a size at the collimator creates very large divergence of the beam, as particle become highly repellent to each other, and would cause downstream beam loss. The Safe ES method was set up to dynamically balance the trade-off between beam size at the collimator and overall beam loss.

The safety of the system is a function of the measured current $I_c$ at the collimator and the performance of the system is a function of a measured current $I_b$ farther downstream. The signals $I_c$ and $I_b$ are the raw readings from sensors. Both signals in their raw form are negative as the $H^{-}$ ion beam has a negative current. If the magnitude of $I_c$ becomes too large, or too negative, then a larger fraction of beam is contacting the collimator leading to lack of safety. On the other hand, if $I_b$ becomes large (more negative) then we have a large fraction of beam surviving at the end of the line - therefore we desire that $I_b$ be minimized as much as possible. Note that in this tuning problem, it is not possible to tune $I_c$ identically to zero. $I_c$ should merely be constrained approximately within some small range, which is why Safe ES is an ideal choice of algorithm in this scenario.

The safety goal was to keep the measured current intercepted at the LEBT collimator, $I_c$, at a value greater than approximately $-0.5$ mA. The reading of $I_c(t)$ lies in the range $[-2.5, 0]$, and we define the safety function defined as 
\begin{equation}
h(t) = w_h (0.5+I_c(t)),
\end{equation}
with $w_h=10$ so that $h\geq0$ corresponds to the region of safety where $-0.5 \leq I_c \leq 0$. We implement a 10 point average of the raw data (shown in Fig. \ref{fig:screenshot_safety_and_error}) to smooth the measurements of $J$ and $h$ - see the smoothed measurements in Fig. \ref{fig:experimental_safety_and_error}. The beam loss was minimized by simply defining $J=I_b$. Minimizing loss in this case is the same as minimizing the beam current because $I_b$ is always negative. The tuning parameters used are the three LEBT steering magnet strengths, given by the channels ``TBSM402P01”, ``TBSM502P01”, and ``TBSM602P01”. The physical locations of the components can be shown in Fig. \ref{fig:lebt_diagram} as ``TBSM04'', ``TBSM05'', and ``TBSM06''. $I_b$ and $I_c$ have channel labels ``TDCM001102'' and ``TDBA001101'' respectively. The Safe ES parameters used are $dt = 0.359$, $a = 0.1$, $k = 0.03$, $\omega_f = 1$, $\omega_1 = 1$, $\omega_2 = 1.375$, $\omega_3 = 1.75$ and $c=1$. 

Results of tuning in the first 100 steps are are shown in Fig. \ref{fig:experimental_safety_and_error} and \ref{fig:experimental_parameters_beamtime}. In Fig. \ref{fig:screenshot_safety_and_error} we show that initially the steering magnets are manually tuned to achieve a desirable operating condition. We then detune the magnets and run the algorithm around minute 4. The algorithm demonstrates it can recover the original operating condition, minimizing $I_b$ to -1.8 mA and driving $I_c$ to near -0.5 mA. The data from Fig. \ref{fig:experimental_safety_and_error} - \ref{fig:experimental_parameters_beamtime} is generated using a 10-point average of the data during the times shown between the dotted lines in Fig. \ref{fig:screenshot_safety_and_error}.

This is an example of multi-variable Safe ES where the parameters of the algorithm converge to the barrier of $h$, achieving practical safety, while also minimizing the objective. Additionally, we have demonstrated that this algorithm can save an accelerator system operating unsafely and drive the system towards a safe operating condition. 

\section{CONCLUSION}
In this paper we demonstrate that Safe ES has a number of uses in tuning various accelerator systems and subsystems. We have shown that in two validated simulations of beamlines at LANSCE, we can use Safe ES to perform tuning. Additionally, we demonstrate that we are able to do this with an in-hardware demonstration of tuning the LEBT section of the accelerator. The advantage of this method is that we require no knowledge or prior gathering of data of the underlying accelerator system and can guarantee practically safe operation. Furthermore, compared to other methods commonly used in particle accelerator systems, Safe ES is very simple to implement.

\balance
\bibliographystyle{IEEEtran}
\bibliography{IEEEabrv,references}

\begin{thebibliography}{10}
\providecommand{\url}[1]{#1}
\csname url@samestyle\endcsname
\providecommand{\newblock}{\relax}
\providecommand{\bibinfo}[2]{#2}
\providecommand{\BIBentrySTDinterwordspacing}{\spaceskip=0pt\relax}
\providecommand{\BIBentryALTinterwordstretchfactor}{4}
\providecommand{\BIBentryALTinterwordspacing}{\spaceskip=\fontdimen2\font plus
\BIBentryALTinterwordstretchfactor\fontdimen3\font minus
  \fontdimen4\font\relax}
\providecommand{\BIBforeignlanguage}[2]{{%
\expandafter\ifx\csname l@#1\endcsname\relax
\typeout{** WARNING: IEEEtran.bst: No hyphenation pattern has been}%
\typeout{** loaded for the language `#1'. Using the pattern for}%
\typeout{** the default language instead.}%
\else
\language=\csname l@#1\endcsname
\fi
#2}}
\providecommand{\BIBdecl}{\relax}
\BIBdecl

\bibitem{king1999800}
N.~King, E.~Ables, K.~Adams, K.~Alrick, J.~Amann, S.~Balzar, P.~Barnes~Jr,
  M.~Crow, S.~Cushing, J.~Eddleman \emph{et~al.}, ``An 800-mev proton
  radiography facility for dynamic experiments,'' \emph{Nuclear Instruments and
  Methods in Physics Research Section A: Accelerators, Spectrometers, Detectors
  and Associated Equipment}, vol. 424, no.~1, pp. 84--91, 1999.

\bibitem{williams2022practically}
A.~Williams, M.~Krsti{\'c}, and A.~Scheinker, ``Practically safe extremum
  seeking,'' in \emph{2022 IEEE 61st Conference on Decision and Control
  (CDC)}.\hskip 1em plus 0.5em minus 0.4em\relax IEEE, 2022, pp. 1993--1998.

\bibitem{williamsCDC2023}
------, ``Semi-global practical extremum seeking with practical safety,'' in
  \emph{2023 IEEE 62st Conference on Decision and Control (CDC)}.\hskip 1em
  plus 0.5em minus 0.4em\relax IEEE, 2023.

\bibitem{krstic2000stability}
M.~Krstic and H.-H. Wang, ``Stability of extremum seeking feedback for general
  nonlinear dynamic systems,'' \emph{Automatica}, vol.~36, no.~4, pp. 595--601,
  2000.

\bibitem{ariyur2003real}
K.~B. Ariyur and M.~Krstic, \emph{Real-time optimization by extremum-seeking
  control}.\hskip 1em plus 0.5em minus 0.4em\relax John Wiley \& Sons, 2003.

\bibitem{ames2016control}
A.~D. Ames, X.~Xu, J.~W. Grizzle, and P.~Tabuada, ``Control barrier function
  based quadratic programs for safety critical systems,'' \emph{IEEE
  Transactions on Automatic Control}, vol.~62, no.~8, pp. 3861--3876, 2016.

\bibitem{nesic2010unifying}
D.~Nesi{\'c}, Y.~Tan, W.~H. Moase, and C.~Manzie, ``A unifying approach to
  extremum seeking: Adaptive schemes based on estimation of derivatives,'' in
  \emph{49th IEEE conference on decision and control (CDC)}.\hskip 1em plus
  0.5em minus 0.4em\relax IEEE, 2010, pp. 4625--4630.

\bibitem{tan2005non}
Y.~Tan, D.~Ne{\v{s}}i{\'c}, and I.~Mareels, ``On non-local stability properties
  of extremum seeking control,'' \emph{IFAC Proceedings Volumes}, vol.~38,
  no.~1, pp. 550--555, 2005.

\bibitem{tan2006non}
\BIBentryALTinterwordspacing
Y.~Tan, D.~Nešić, and I.~Mareels, ``On non-local stability properties of
  extremum seeking control,'' \emph{Automatica}, vol.~42, no.~6, pp. 889--903,
  2006. [Online]. Available:
  \url{https://www.sciencedirect.com/science/article/pii/S0005109806000690}
\BIBentrySTDinterwordspacing

\bibitem{guay2015constrained}
M.~Guay, E.~Moshksar, and D.~Dochain, ``A constrained extremum-seeking control
  approach,'' \emph{International Journal of Robust and Nonlinear Control},
  vol.~25, no.~16, pp. 3132--3153, 2015.

\bibitem{labar2019constrained}
C.~Labar, E.~Garone, M.~Kinnaert, and C.~Ebenbauer, ``Constrained extremum
  seeking: a modified-barrier function approach,'' \emph{IFAC-PapersOnLine},
  vol.~52, no.~16, pp. 694--699, 2019.

\bibitem{liao2019constrained}
C.-K. Liao, C.~Manzie, A.~Chapman, and T.~Alpcan, ``Constrained extremum
  seeking of a mimo dynamic system,'' \emph{Automatica}, vol. 108, p. 108496,
  2019.

\bibitem{poveda2015shahshahani}
J.~I. Poveda and N.~Quijano, ``Shahshahani gradient-like extremum seeking,''
  \emph{Automatica}, vol.~58, pp. 51--59, 2015.

\bibitem{wang2022extremum}
Z.~Wang, X.~Zhou, and J.~Wang, ``Extremum-seeking-based adaptive model-free
  control and its application to automated vehicle path tracking,''
  \emph{IEEE/ASME Transactions on Mechatronics}, vol.~27, no.~5, pp.
  3874--3884, 2022.

\bibitem{ye2020extremum}
M.~Ye, G.~Hu, and S.~Xu, ``An extremum seeking-based approach for nash
  equilibrium seeking in n-cluster noncooperative games,'' \emph{Automatica},
  vol. 114, p. 108815, 2020.

\bibitem{paz2019model}
P.~Paz, T.~R. Oliveira, A.~V. Pino, and A.~P. Fontana, ``Model-free
  neuromuscular electrical stimulation by stochastic extremum seeking,''
  \emph{IEEE transactions on control systems technology}, vol.~28, no.~1, pp.
  238--253, 2019.

\bibitem{kumar2019extremum}
S.~Kumar, A.~Mohammadi, D.~Quintero, S.~Rezazadeh, N.~Gans, and R.~D. Gregg,
  ``Extremum seeking control for model-free auto-tuning of powered prosthetic
  legs,'' \emph{IEEE Transactions on Control Systems Technology}, vol.~28,
  no.~6, pp. 2120--2135, 2019.

\bibitem{huang2018robust}
X.~Huang, ``Robust simplex algorithm for online optimization,'' \emph{Physical
  Review Accelerators and Beams}, vol.~21, no.~10, p. 104601, 2018.

\bibitem{aiba2012ultra}
M.~Aiba, M.~B{\"o}ge, N.~Milas, and A.~Streun, ``Ultra low vertical emittance
  at sls through systematic and random optimization,'' \emph{Nuclear
  Instruments and Methods in Physics Research Section A: Accelerators,
  Spectrometers, Detectors and Associated Equipment}, vol. 694, pp. 133--139,
  2012.

\bibitem{schuster2007beam}
E.~Schuster, C.~Xu, N.~Torres, E.~Morinaga, C.~Allen, and M.~Krstic, ``Beam
  matching adaptive control via extremum seeking,'' \emph{Nuclear Instruments
  and Methods in Physics Research Section A: Accelerators, Spectrometers,
  Detectors and Associated Equipment}, vol. 581, no.~3, pp. 799--815, 2007.

\bibitem{modelscheinker}
\BIBentryALTinterwordspacing
A.~Scheinker, X.~Pang, and L.~Rybarcyk, ``Model-independent particle
  accelerator tuning,'' \emph{Phys. Rev. ST Accel. Beams}, vol.~16, p. 102803,
  Oct 2013. [Online]. Available:
  \url{https://link.aps.org/doi/10.1103/PhysRevSTAB.16.102803}
\BIBentrySTDinterwordspacing

\bibitem{scheinker2019model}
A.~Scheinker, D.~Bohler, S.~Tomin, R.~Kammering, I.~Zagorodnov, H.~Schlarb,
  M.~Scholz, B.~Beutner, and W.~Decking, ``Model-independent tuning for
  maximizing free electron laser pulse energy,'' \emph{Physical review
  accelerators and beams}, vol.~22, no.~8, p. 082802, 2019.

\bibitem{7859370}
A.~Scheinker, X.~Huang, and J.~Wu, ``Minimization of betatron oscillations of
  electron beam injected into a time-varying lattice via extremum seeking,''
  \emph{IEEE Transactions on Control Systems Technology}, vol.~26, no.~1, pp.
  336--343, 2018.

\bibitem{scheinker2020onlineMO}
A.~Scheinker, S.~Hirlaender, F.~M. Velotti, S.~Gessner, G.~Z. Della~Porta,
  V.~Kain, B.~Goddard, and R.~Ramjiawan, ``Online multi-objective particle
  accelerator optimization of the awake electron beam line for simultaneous
  emittance and orbit control,'' \emph{AIP Advances}, vol.~10, no.~5, 2020.

\bibitem{scheinker2021extremum}
A.~Scheinker, E.-C. Huang, and C.~Taylor, ``Extremum seeking-based control
  system for particle accelerator beam loss minimization,'' \emph{IEEE
  Transactions on Control Systems Technology}, vol.~30, no.~5, pp. 2261--2268,
  2021.

\bibitem{arpaia2021machine}
P.~Arpaia, G.~Azzopardi, F.~Blanc, G.~Bregliozzi, X.~Buffat, L.~Coyle, E.~Fol,
  F.~Giordano, M.~Giovannozzi, T.~Pieloni \emph{et~al.}, ``Machine learning for
  beam dynamics studies at the cern large hadron collider,'' \emph{Nuclear
  Instruments and Methods in Physics Research Section A: Accelerators,
  Spectrometers, Detectors and Associated Equipment}, vol. 985, p. 164652,
  2021.

\bibitem{kirschner2022tuning}
J.~Kirschner, M.~Mutn{\`y}, A.~Krause, J.~C. de~Portugal, N.~Hiller, and
  J.~Snuverink, ``Tuning particle accelerators with safety constraints using
  bayesian optimization,'' \emph{Physical Review Accelerators and Beams},
  vol.~25, no.~6, p. 062802, 2022.

\bibitem{kirschner2019adaptive}
J.~Kirschner, M.~Mutny, N.~Hiller, R.~Ischebeck, and A.~Krause, ``Adaptive and
  safe bayesian optimization in high dimensions via one-dimensional
  subspaces,'' in \emph{International Conference on Machine Learning}.\hskip
  1em plus 0.5em minus 0.4em\relax PMLR, 2019, pp. 3429--3438.

\bibitem{duris2020bayesian}
J.~Duris, D.~Kennedy, A.~Hanuka, J.~Shtalenkova, A.~Edelen, P.~Baxevanis,
  A.~Egger, T.~Cope, M.~McIntire, S.~Ermon \emph{et~al.}, ``Bayesian
  optimization of a free-electron laser,'' \emph{Physical review letters}, vol.
  124, no.~12, p. 124801, 2020.

\bibitem{john2021real}
J.~S. John, C.~Herwig, D.~Kafkes, J.~Mitrevski, W.~A. Pellico, G.~N. Perdue,
  A.~Quintero-Parra, B.~A. Schupbach, K.~Seiya, N.~Tran \emph{et~al.},
  ``Real-time artificial intelligence for accelerator control: A study at the
  fermilab booster,'' \emph{Physical Review Accelerators and Beams}, vol.~24,
  no.~10, p. 104601, 2021.

\bibitem{edelen2020machine}
A.~Edelen, N.~Neveu, M.~Frey, Y.~Huber, C.~Mayes, and A.~Adelmann, ``Machine
  learning for orders of magnitude speedup in multiobjective optimization of
  particle accelerator systems,'' \emph{Physical Review Accelerators and
  Beams}, vol.~23, no.~4, p. 044601, 2020.

\bibitem{blokland2022uncertainty}
W.~Blokland, K.~Rajput, M.~Schram, T.~Jeske, P.~Ramuhalli, C.~Peters,
  Y.~Yucesan, and A.~Zhukov, ``Uncertainty aware anomaly detection to predict
  errant beam pulses in the oak ridge spallation neutron source accelerator,''
  \emph{Physical Review Accelerators and Beams}, vol.~25, no.~12, p. 122802,
  2022.

\bibitem{wolski2022transverse}
A.~Wolski, M.~A. Johnson, M.~King, B.~L. Militsyn, and P.~H. Williams,
  ``Transverse phase space tomography in an accelerator test facility using
  image compression and machine learning,'' \emph{Physical Review Accelerators
  and Beams}, vol.~25, no.~12, p. 122803, 2022.

\bibitem{mayet2022predicting}
F.~Mayet, M.~Hachmann, K.~Floettmann, F.~Burkart, H.~Dinter, W.~Kuropka,
  T.~Vinatier, and R.~Assmann, ``Predicting the transverse emittance of space
  charge dominated beams using the phase advance scan technique and a fully
  connected neural network,'' \emph{Physical Review Accelerators and Beams},
  vol.~25, no.~9, p. 094601, 2022.

\bibitem{zhu2021high}
J.~Zhu, Y.~Chen, F.~Brinker, W.~Decking, S.~Tomin, and H.~Schlarb,
  ``High-fidelity prediction of megapixel longitudinal phase-space images of
  electron beams using encoder-decoder neural networks,'' \emph{Physical Review
  Applied}, vol.~16, no.~2, p. 024005, 2021.

\bibitem{hirlaender2020model}
S.~Hirlaender and N.~Bruchon, ``Model-free and bayesian ensembling model-based
  deep reinforcement learning for particle accelerator control demonstrated on
  the fermi fel,'' \emph{arXiv preprint arXiv:2012.09737}, 2020.

\bibitem{kain2020sample}
V.~Kain, S.~Hirlander, B.~Goddard, F.~M. Velotti, G.~Z. Della~Porta,
  N.~Bruchon, and G.~Valentino, ``Sample-efficient reinforcement learning for
  cern accelerator control,'' \emph{Physical Review Accelerators and Beams},
  vol.~23, no.~12, p. 124801, 2020.

\bibitem{scheinker2018demonstration}
A.~Scheinker, A.~Edelen, D.~Bohler, C.~Emma, and A.~Lutman, ``Demonstration of
  model-independent control of the longitudinal phase space of electron beams
  in the linac-coherent light source with femtosecond resolution,''
  \emph{Physical review letters}, vol. 121, no.~4, p. 044801, 2018.

\bibitem{scheinker2021adaptiveML}
A.~Scheinker, ``Adaptive machine learning for time-varying systems: low
  dimensional latent space tuning,'' \emph{Journal of Instrumentation},
  vol.~16, no.~10, p. P10008, 2021.

\bibitem{scheinker2023adaptiveML}
A.~Scheinker, F.~Cropp, and D.~Filippetto, ``Adaptive autoencoder latent space
  tuning for more robust machine learning beyond the training set for
  six-dimensional phase space diagnostics of a time-varying ultrafast
  electron-diffraction compact accelerator,'' \emph{Physical Review E}, vol.
  107, no.~4, p. 045302, 2023.

\bibitem{AmesAutomotive}
A.~D. Ames, X.~Xu, J.~W. Grizzle, and P.~Tabuada, ``Control barrier function
  based quadratic programs for safety critical systems,'' \emph{IEEE
  Transactions on Automatic Control}, vol.~62, pp. 3861--3876, 2017.

\bibitem{AmesCruiseControl}
A.~D. Ames, J.~W. Grizzle, and P.~Tabuada, ``Control barrier function based
  quadratic programs with application to adaptive cruise control,'' in
  \emph{IEEE Conference on Decision and Control}, 2014, pp. 6271--6278.

\bibitem{scheinker2014extremum}
A.~Scheinker and M.~Krsti{\'c}, ``Extremum seeking with bounded update rates,''
  \emph{Systems \& Control Letters}, vol.~63, pp. 25--31, 2014.

\bibitem{Roy_2018}
\BIBentryALTinterwordspacing
P.~K. Roy, C.~E. Taylor, C.~Pillai, and Y.~K. Batygin, ``Comparison of profile
  measurements and transport beam envelope predictions along the 80-m lansce
  prad beamline,'' \emph{Journal of Physics: Conference Series}, vol. 1067,
  no.~6, p. 062002, sep 2018. [Online]. Available:
  \url{https://dx.doi.org/10.1088/1742-6596/1067/6/062002}
\BIBentrySTDinterwordspacing

\bibitem{pang2015advances}
X.~Pang \emph{et~al.}, ``Advances in proton linac online modeling,''
  \emph{IPAC’15}, pp. 2423--2427, 2015.

\bibitem{huang2021enhanced}
E.-C. Huang, C.~E. Taylor, and P.~M. Anisimov, ``Enhanced beam diagnostics with
  existing bppms via gpu-powered multi-particle simulation,'' Los Alamos
  National Lab.(LANL), Los Alamos, NM (United States), Tech. Rep., 2021.

\end{thebibliography}

\end{document}